\documentclass[11pt,twoside]{article}

\usepackage{asp2006}
\usepackage{epsf}
\usepackage{psfig}
\usepackage{lscape}

\begin{document}
\title{\textit{Swift} Observations of Shock Evolution in RS Ophiuchi}   
\author{M.~F.~Bode,${^1}$ J.~P.~Osborne,${^2}$ K.~L.~Page,${^2}$ A.~P.~Beardmore,${^2}$
T.~J.~O'Brien,$^3$ J-U.~Ness,${^4}$ S.~Starrfield,${^4}$ G.~K.~Skinner,${^5}$
M.~J.~Darnley,${^1}$ J.~J.~Drake,${^6}$ A.~Evans,${^7}$ S.~P.~S.~Eyres,${^8}$
J.~Krautter,${^9}$ and G.~Schwarz$^{10}$}

\affil{$^1$~Astrophysics Research Institute, Liverpool John Moores Univ., UK\\ 
$^2$~Department of Physics and Astronomy, University of Leicester, UK\\ 
$^3$~School of Physics and Astronomy, University of Manchester, UK\\ 
$^4$~School of Earth and Space Exploration, Arizona State Univ., USA\\ 
$^5$~NASA Goddard Space Flight Center, Greenbelt, MD, USA\\ 
$^6$~Harvard-Smithsonian Center for Astrophysics, Cambridge, MA, USA\\ 
$^7$~Astrophysics Group, Keele University, ST5 5BG, UK\\
$^8$~Centre for Astrophysics, University of Central Lancashire, UK\\ 
$^9$~Landessternwarte, Heidelberg, Germany\\ 
$^{10}$~Department of Geology and Astronomy, West Chester Univ., USA}


\vspace{3mm}

\vspace{-\baselineskip}

\begin{abstract} 
Our \textit{Swift} observations of RS Oph form an unprecedented X-ray dataset to undertake
investigations of both the central source and the interaction of the outburst ejecta with the
circumstellar environment. Over the first month, the XRT data are dominated by emission from rapidly
evolving shocks. We discuss the differences in derived parameters from those found for \textit{RXTE} at
early times and the evolution of the X-ray emission to much later times. It is apparent that at late
times several emission components are present. We find no strong evidence of the proposed shock
break-out in our data.
\end{abstract}

\section{Introduction}  
Prior to 1985, evidence for the interaction of outburst ejecta with the circumstellar environment
in RS~Oph came from observations of coronal lines in optical spectra, the narrowing of initially
broad emission lines, and the evolution of superimposed narrow emission features
\citep[see e.g.][and references therein]{mas87_bs}. \textit{EXOSAT} observations from $t = 55$ to 251 days post-outburst in
1985 showed that RS Oph was a bright, but rapidly declining, X-ray source
\citep{mas87_bs}. Bode \& Kahn (1985) formulated an analytical model 
and concluded that RS Oph evolved like a supernova remnant, but on timescales around $10^5$ times
faster. Subsequently, O'Brien et al. (1992) developed detailed hydrodynamical models of the interaction of the
ejecta with the red giant wind to derive parameters of the explosion and the circumstellar environment.

\section{The First Month}

Following notification of the latest outburst, our pointed \textit{Swift} observations began 3.17 days
after optical discovery \citep{bod06_bs}. The source was already very bright in the XRT band ($14.2 \pm
0.2$ c.p.s.) and had further increased in flux to $31.5 \pm 0.2$ c.p.s. by the time of the next
observation at $t = 5.03$ days. It then began a gradual decline in flux which continued until $t = 29$
days. 

\begin{figure}[!ht]
\plotfiddle{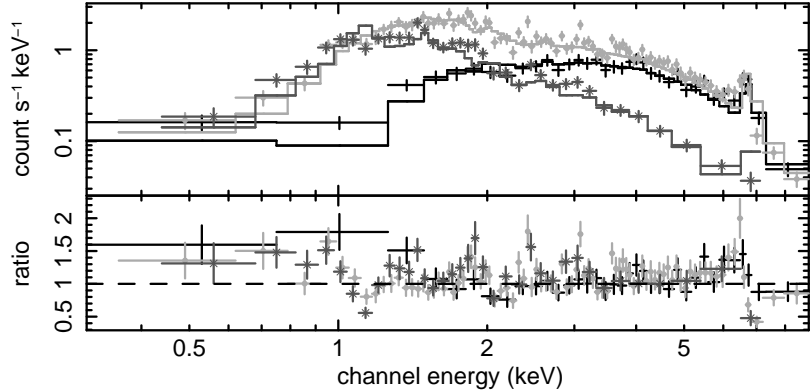}{45mm}{270}{48}{48}{-185}{263}
\caption{\textit{Swift} XRT data on days 3.17 (black), 5.05 (grey) and 13.6 (black stars) compared in each case to a single temperature \textit{mekal} model spectral fit with decreasing absorption (see text and Bode et al. (2006) for further details).}\label{fig1}
\end{figure}

During this period, the spectrum was obviously that of an emission line source superimposed on a broad
continuum (see Figure~\ref{fig1}, also Ness et al. and Nelson et al., these proceedings) 
which was then fitted with a high temperature thermal plasma (\textit{mekal}) model. A wide range of
parameter space was explored using combinations of multi-temperature components, varying abundances and
absorbing columns. Ultimately it was decided to use a simplistic, first order, approach. This involved
single temperature, solar abundance, fits to the data. The column was taken to comprise two components;
the interstellar absorption fixed at $2.4 \times 10^{21}$ cm$^{-2}$ throughout \citep{hje86_bs}, but with
an additional column for the red giant wind overlying the shock being a free parameter. The distance to
RS Oph was assumed to be 1.6 kpc. Subsequent investigation of the BAT data showed that the source was
also detected at outburst in the lowest energy channel (14--25~keV, and possibly in the 25--50~keV as
well -- see Skinner et al. and Senziani et al., these proceedings). Extrapolation of the fits to the XRT data showed the BAT
emission was consistent with these \textit{mekal} model fits.

\begin{figure}[!ht]
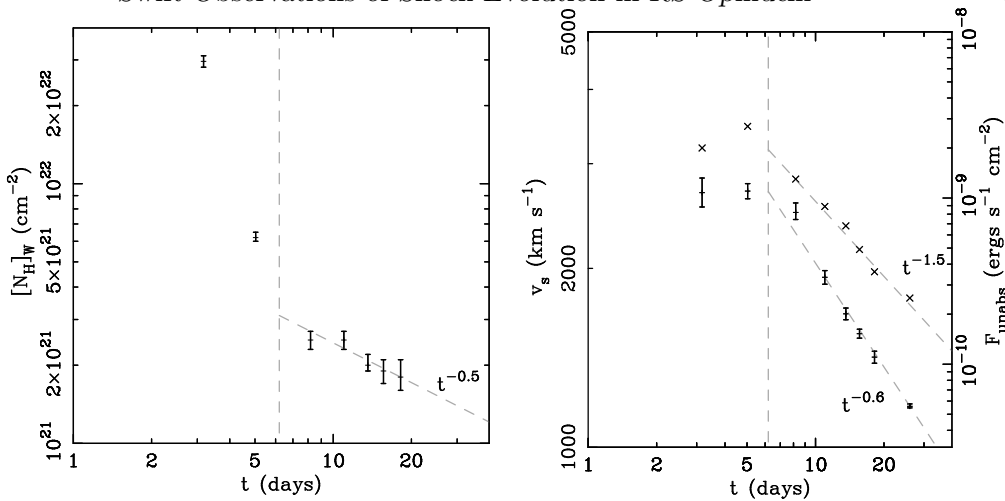

\plotfiddle{bode_fig2a.eps}{45mm}{270}{34}{34}{-202}{166}
\plotfiddle{bode_fig2b.eps}{0mm}{270}{34}{34}{-8}{189}
\caption{Evolution of the overlying wind column (left panel) and forward shock velocity and unabsorbed
flux (crosses, both right panel) derived from spectral fits (see text and \cite{bod06_bs} for further
details)}\label{fig2}
\end{figure}

From Figure~\ref{fig2}, it is apparent that from around 6 days after the outburst, the resulting
parameters decline as a simple power-law with time. As noted in \cite{bod06_bs}, this behaviour is more
consistent with that expected in Phase III of remnant evolution than that in Phase II, contrary to the
conclusions in Bode \& Kahn (1985), or the behaviour of the velocity derived by Sokoloski et al. (2006) from fits to the
early \textit{RXTE} data. However, we note that the later velocities derived from \textit{RXTE} data do
seem to fall away more steeply and that the fits to the derived fluxes show a behaviour more akin to
that expected in Phase III \citep[][see also Tatischeff and Hernanz, these proceedings]{sok06_bs}. 

Attention has been drawn to apparent discrepancies between fits to the XRT and \textit{RXTE} data at
early times. For the first epoch, \textit{RXTE} and \textit{Swift} observations were taken only 2 hr 20
min apart. Our fits to the combined data set found no significant changes from our initial results for
the XRT alone. All subsequent data were obtained at temporally more separated epochs. It should be
noted that the \textit{mekal} models include both line and continuum emission, whereas the
\textit{RXTE} results were fitted with a thermal bremsstrahlung model, plus a superimposed  Fe line,
only (and also for a fixed value of the absorbing column). Other differences apart, one possibility is
therefore that at early times, the electrons (giving rise to the continuum) and the ions (giving the
lines) are not in thermal equilibrium, naturally explaining the differing results of the two fitting
techniques. The effects of source geometry and varying abundances also need to be explored.

A change in behaviour seen in Figure~\ref{fig2} at around $t = 6$ days is also apparent in the fluxes
derived from the \textit{RXTE} fits \citep{sok06_bs}. Bode et al. (2006) identified this with the end of Phase I
of remnant evolution. Vaytet et al. (these proceedings) have modelled this in terms of the end of
significant mass loss in a fast wind from the WD \citep[see also][]{vay07_bs}.
Das et al. (2006) also find a
change from roughly constant to power-law decline in velocities derived from infrared spectroscopy
after $t = 4$ days, and with a subsequent slope more consistent with Phase III than Phase II behaviour.

\section{Late Time Evolution and Future Work}

\begin{figure}[!ht]
\plotfiddle{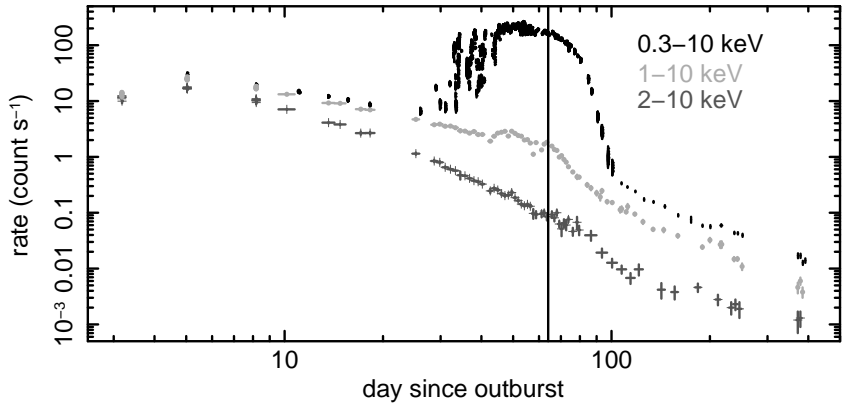}{46mm}{270}{50}{50}{-200}{270}
\caption{XRT count rate vs time in three energy bands. The vertical line indicates an abrupt change in slope in the 1-10 keV count rate at $t = 64$ days.}\label{fig3}
\end{figure}

After the first month, the Super Soft Source came to dominate the spectrum 
and this meant that meaningful fits to any shock component could not easily
be undertaken. 
At $t \sim 60$ days, the SSS began to decline and this phase was thought to have ended at around 90
days. To explore the evolution in more detail, we split the source flux into different energy bands
(see Figure~\ref{fig3}). This clearly shows that while there was a dramatic change in the overall count
rate at the time of the emergence of the SSS, taking energies above 1 keV, the gradual decline
continued. There was however a change in slope evident in the 1-10 keV range after $t = 64$ days which
is not as clear at higher energies. We note that the 35s modulation in SSS flux, thought to arise from
processes on the WD surface, also ceased to be detectable at $t = 65.9$ days (Beardmore et al., these
proceedings). The break at $t = 64$ days is thus most likely due to a change in emission from the
central source, rather than the shocked wind and there is no evidence therefore of ``shock break-out''.
Indeed, taking a wind velocity of 20 km s$^{-1}$ blowing for the 21 year inter-outburst period with the
results for the evolution of shock velocity from  our fits, and shock radius at $t = 14$ days from
O'Brien et al. (2006), shock break-out would not be expected to occur before $\sim1300$ days and thus may not be
observable at all.

We have undertaken spectral fitting of a range of epochs between 90 and 391 days. In all cases with
sufficient source counts it appears there are at least two distinct spectral components present. We may
need to perform adequate modelling of the SSS even at these times to remove this before modelling the
remaining emission from the shocked wind. More realistic hydrodynamic models are being developed,
including cooling \citep{vay07_bs} and 3-D modelling to take into account remnant structure revealed by
VLBI and HST imaging \citep[see O'Brien et al. and Harman et al., these proceedings, and][]{rup08_bs}.

\acknowledgements The authors are grateful to the \textit{Swift} Mission Operations Center staff for
their superb support of the observations reported here.



\end{document}